\def \bc {\begin{center}}
\def \ec {\end{center}}

\def \bfr {\begin{flushright}}
\def \efr {\end{flushright}}

\def \v {\vskip}

\def \ba {\begin{array}}
\def \ea {\end{array}}

\def \bea {\begin{eqnarray}}
\def \eea {\end{eqnarray}}

\def \be {\begin{equation}}
\def \ee {\end{equation}}
\def \p {\partial}

\def \d {\hbox{d}\,}
\def \square {\hbox{$\sqcup\!\!\!\!\sqcap$}} 

\documentstyle[a4,12pt]{article}

\begin{document}
%
\thispagestyle{empty}


\v.5cm

\v5mm
\begin{center}

{\bf RELATIVISTIC FIELD EQUATIONS FROM HIGHER-ORDER 
POLARIZATIONS OF THE POINCAR\'E GROUP}
\v5mm

{\it  Miguel Navarro$^{1,2}$\footnote{http://www.ugr.es/$\tilde{}$mnavarro; 
e-mail: mnavarro@ugr.es },
Manuel Calixto$^{1,3}$\footnote{e-mail: calixto@ugr.es}  
and
V\'\i ctor Aldaya$^{1,4}$\footnote{e-mail: valdaya@ugr.es}
}

\v3mm
\end{center}

\begin{enumerate}

\item Instituto Carlos I de F\'\i sica Te\'orica y Computacional,
Facultad  de  Ciencias, Universidad de Granada, Campus de Fuentenueva,
18002, Granada, Spain.

\item Instituto de Matem\'aticas y F\'\i sica 
Fundamental, CSIC, Serrano 113-123, 28006 Madrid, Spain. 

\item Departamento de F\'\i sica Te\'orica y del Cosmos. 
Facultad de Ciencias. Campus de Fuentenueva, 18002. Granada. Spain. 

\item IFIC, Centro Mixto Universidad de
Valencia-CSIC, Burjassot 46100-Valencia, Spain.

\end{enumerate} 
\v2mm

\centerline{Abstract}
\v2mm
The theory of free relativistic fields is shown to arise 
in a unified manner from higher-order, 
configuration-space, irreducible    
representations of the Poincar\'e group. 
A de Sitter subalgebra, in the massive case, and a Poincar\'e
subalgebra, in the massless case, of the enveloping 
algebra of the Poincar\'e 
group are the suitable higher-order polarizations. 
In particular, a simple group-theoretic derivation of the Dirac 
equation is given. 

\v2mm
\noindent PAC numbers: 03.65.Pm, 03.50.-z, 11.30.Cp

\newpage 
\setcounter{page}{1}

\section{Introduction}

The theory of relativistic field equations is older than 
the Special Theory of Relativity. In fact, 
it was mainly  the Poincar\'e 
invariance of Maxwell equations for electromagnetism 
-- equations which had been obtained by Maxwell 
from theoretical considerations upon previous work by 
Coulomb, Faraday and Ampere -- 
that motivated the shift from the Galileo group 
to the Poincar\'e group as the relativity group of Physics. 
However, it was the quantum revolution of the twenties that actually 
impelled this subject to acquire the relevance it has gained then 
since.  

The discovery and, in general, the 
study of the relativistic field equations has followed 
basically two approaches, Dirac's and Wigner's. In Dirac's 
approach,  the equations are postulated firstly and their invariance 
under the Poincar\'e group is ``discovered'' afterwards. 
In Wigner's approach, on the contrary, 
the representations of the Poincar\'e group 
are calculated first and then it is shown that 
the space which supports these representations equals 
the space of solutions of some (relativistic) field equations. 
These approaches have, no doubt, been very fruitful. However,  
both are somewhat unsatisfactory as in neither of the two the 
relativistic field equations are directly derived from the relativity 
group and directly in configuration space. 
As a consequence of this, the actual group-theoretic 
origin of the Dirac equation remains unclear. 
Also, group-theoretic treatments of 
the relativistic field equations have, in general,  
fail to enter the standard 
literature on Quantum Field Theory, which is regrettable given 
the relevance of the subject 
(a remarkable exception to this rule is Ref. \cite{[Weinberg]}).   
This and other shortcomings of Wigner's (and Dirac's) approaches 
may be regarded as symptoms that the theory of 
relativistic equations is in need of a process
of {\it aggiornamento}, which, by using modern representation 
techniques should clarify, simplifly and thus improve it. 

The present paper is meant to be a first step in this process of 
{\it aggiornamento}. We show how the 
relativistic field equations can be obtained -- in configuration space   
and directly from the relativity group -- as providing linear,  
finite-component, irreducible 
representations of the Poincar\'e group. 
We basically use standard techniques of the theory of 
quantization on a coset space $G/H$,  but generalize them so as to fetch   
some higher-order polarization techniques 
from the Group Approach to Quantization 
(GAQ) formalism (see Ref. \cite{[marmo]} and references therein).  
These higher-order polarization conditions, which are being laid on 
solid grounds within the GAQ formalism 
\cite{[marmo],[higherpjmp],[higherpcmp]} generalize both 
the familiar Casimir-operator conditions and the formalism 
of induced representation, which uses first-order polarizations only. 

The main goal of the present paper is to show how these higher-order 
quantizations give rise to the right configuration-space 
equations of motion for the classical fields.  
In this way, we provide a direct group-theoretic construction of
these equations, in particular Dirac equation. 
As far as we know, no derivation similar to ours  
has been presented before.  

The structure of the present paper is as follows.  
The mathematical foundations of our approach are presented in Sect. 2 
and are applied in Sect. 3 to the Poincar\'e group.  
In Sect. 4, we touch upon several facts which may be useful 
to derive maximum benefit from our study. 
The Appendix, which contents some general features 
of the Clifford and Kemmel algebras, 
has been added to facilitate  
the reading of the present paper. 

Excellent complementary discussions on the 
subject as well as more references to original works  
can be found in Refs. \cite{[Weinberg]} and \cite{[corson]}-\cite{[velo]}. 

\section{Higher-order polarizations}

Our developments are based on a result which has recently  
been proven in the context of GAQ (see Ref. \cite{[marmo]}). 
A version of this statement, suitably taylored for the subject at hand,  
can be presented in the following form: 

Let $G$ be a connected and simply connected Lie group 
and ${\cal U}^L(G)$ its universal left-enveloping algebra.  
Let $H$ be an Abelian subgroup of $G$ and ${\cal A}$ a 
maximal subalgebra of ${\cal U}^L(G)$ which does not include the
Lie algebra of $H$. Then, $G$ is pseudo-irreducibly represented on the
functions  $\Psi:\>G\longrightarrow C$, which fulfil  

\be Y.\Psi=0,\quad \forall \>Y\in {\cal A}\ee 

Pseudo-irreducibility means here that the Hilbert space is such that 
any differential or pseudodifferential operator commuting with the 
representation is a multiple of the identity, but it may possibly contain  
invariant subspaces which can (only) be distinguished under the action  
of some non pseudo-differential operators which are external 
to the group. 

In other words, the theorem says that by 
imposing a sufficient number of (first- or higher-order) 
left equations of motion we arrive at a representation of the group which 
is basically irreducible. If there appear non pseudo-differential operators 
which mix subspaces which are invariant under the group, 
they can be taken care of, 
case by case, at the end of the procedure  
(for more details and examples see Ref. \cite{[marmo]}). 

This theorem can be extended -- by means of a corollary which is
presented next and which can be 
proved through similar steps to those of the theorem above  
-- to the case we consider in the present paper, in which 
the left-subalgebra ${\cal A}$ is finitely generated 
and is non-trivially represented.  
A  definition will prove useful: 

\noindent{\it Definition}: We shall say that ${\cal B}\subset 
{\cal U}^L(G)$ is a weak subalgebra if it closes with structure constants 
which may involve Casimir operators. 
In other words, ${\cal B}$ closes as a subalgebra if the
Casimir operators are considered to be numbers. 

\noindent{\it Corollary}: Under the same hypothesis of the theorem,  
consider now that ${\cal A}$ is generated by the Casimir operators  
$C_a,\> a=1,...k$, and   
a weak subalgebra ${\cal B}$ which is finite-dimensional  
with basis $\{Y_i\}_{(i=1,...n)}$.
Let $V$ be a finite-dimensional space where 
$\{Y^L_i\}$ is irreducibly represented by $\{\alpha_i\}$, 
where $\alpha_i$ are finite-dimensional matrices. Let 
${\cal H}$ be the space of functions 
$\Psi:\>G \longrightarrow V$ such that 

\bea C_a.\Psi&=&c_a\Psi, \quad a=1,...k \nonumber\\
 Y^L_i.\Psi &=& \alpha_i\Psi, \quad i=1,...n \label{prop10}\eea 
for some numbers $c_a$. Then $G$ is pseudo-irreducibly 
represented on ${\cal H}$.  

The theorem and its corollary does not directly apply to 
the Poincar\'e group, as it is not connected. However, in this case,   
the discrete symmetries --parity $P$, time reversal $T$ and the
product $PT$--  which cause the group not to be connected, and which 
behave much in the same way as the abovementioned non pseudo-differential 
operators which are external to the group,  
turn out to be simple to manage.  
Let us assume we have been able to find out the finite-component 
irreducible representations of the connected part of the
Poincar\'e group, which the theorem above shows us how to achieve.  
Then, a discrete symmetry $I$ either preserves a given representation 
$D$ or changes it to another representation $D^*$. 
In any case, since $I^2=1_{\cal P}$, we must
have, save for a factor, $D^{**}= D$ and any discrete symmetry 
mixes two representations at most. Therefore, the irreducible
representations of the complete group prove to be a discrete sum of
representations of its connected-to-the-identity component. 
More detailed discussions can be found  
in Ref. \cite{[fonda],[naimark],[velo]}. 

\section{The Poincar\'e group and relativistic wave equations}

The Poincar\'e group ${\cal P}$ is the semidirect product of the 
Lorentz group ${\cal L}$ and the group of translations ${\cal T}$: 

\bea \Lambda''&=&\Lambda'\Lambda \\
a''&=&a' +\Lambda'a \label{Poincare}\eea 
With a convenient parameterization of the group, 
the left- and right-invariant vector fields takes the form  
(see Refs. \cite{[groupfields],[annals]}): 

\bea J^L_{\mu\nu}\equiv X^L_{\epsilon_{\mu\nu}}&=& 
{}^LT^{\alpha\beta}_{\mu\nu}\frac{\partial}
{\partial\epsilon^{\alpha\beta}}\nonumber\\
P^L_\mu = X^L_{a_\mu} &=& 
\Lambda_{\alpha\mu}\frac{\partial}{\partial a_\alpha}\label{leftlorentz}\eea
and 

\bea J^R_{\mu\nu}\equiv X^L_{\epsilon_{\mu\nu}}&=& 
{}^RT^{\alpha\beta}_{\mu\nu}\frac{\partial}{\partial\epsilon^{\alpha\beta}} 
+ (\delta^\alpha_\mu\delta_\nu^\beta-\delta_\mu^\beta\delta_\nu^\alpha)\,
a_\beta\frac{\partial}{\partial a^\alpha}\nonumber\\
P^R_\mu = X^R_{a_\mu} &=& 
\frac{\partial}{\partial a_\mu}\label{rightlorentz}\eea 
where ${}^LT^{\alpha\beta}_{\mu\nu}$ and 
${}^RT^{\alpha\beta}_{\mu\nu}$ are functions 
of the parameters $\epsilon^{\alpha\beta}$ of the group the actual 
expression of which we shall not need (the interested reader is referred, 
nonetheless,  to Refs. \cite{[annals],[groupfields]}).  

These vector fields close the familiar algebra: 

\bea \left[P^L_\mu, P^L_\nu\right]&=&0, 
\qquad \left[J^L_{\mu\nu},P^L_\rho\right]=
\eta_{\nu\rho}P^L_\mu -\eta_{\mu\rho}P^L_\nu\>\nonumber\\ 
\left[J^L_{\mu\nu},J^L_{\alpha\beta}\right]&=&  
\eta_{\nu\alpha}J^L_{\mu\beta}-
\eta_{\mu\alpha}J^L_{\nu\beta}+\eta_{\mu\beta}J^L_{\nu\alpha}-
\eta_{\nu\beta}J^L_{\mu\alpha}\label{poincarealg}\eea 

We have ${\cal T}={\cal P}/{\cal L}$ and 
the Minkowski space can be identified with ${\cal T}$. 
Therefore, we may obtain fields in configuration space by 
quantizing the Poincar\'e group with the Lorentz ($\sim SL(2,C)$) 
subgroup as (part of) the polarization. 

Let us consider, therefore, a representation $S$   
of  $SL(2,C)$ which is defined on a finite-dimensional vector space $V$.
For any function $\Psi: {\cal P}\longrightarrow V$, 
$\Psi=\Psi(\Lambda, a)$,  
the Lorentz polarization condition takes the form:

\be  \Psi(\Lambda\Lambda', a)= S(\Lambda'^{-1})\Psi(\Lambda,a) 
\label{plorentz1}\ee 
By taking $\Lambda'=\Lambda^{-1}$, we obtain: 

\be \Psi(\Lambda,a)=
S^{-1}(\Lambda)\Psi(I,a)\equiv S^{-1}(\Lambda)\Phi(a)
\label{plorentz2}\ee 
Therefore, as a desired result of our procedure,  
we have obtained functions which are defined solely 
over Minkowsky space. Moreover,  
Eqs. (\ref{plorentz1}-\ref{plorentz2}) and the natural (left) action of
the group on the functions $\Psi$ imply that 
the Poincar\'e group acts on 
the physical  fields $\Phi$ as desired: 

\be \left((\Lambda,a)\Phi\right)(x) = 
S(\Lambda)\Phi(\Lambda^{-1}(x-a))\label{poincare3}\ee

Eq. (\ref{plorentz1}) means that the left-invariant 
``angular-momentum'' 
vector fields $J^{L}_{\mu\nu}$ have been represented by 
a finite-dimensional matrix $-S_{\mu\nu}$ 

\be J^L_{\mu\nu}\rightarrow -S_{\mu\nu}\nonumber\ee  
where

\be S(\Lambda)=\exp\{\frac12\epsilon^{\mu\nu}S_{\mu\nu}\}
\label{sigma}\ee 
On the other hand, Eq. (\ref{poincare3}) means that the 
physical ``angular-momentum'' operators $J_{\mu\nu}\equiv
J^{R}_{\mu\nu}$ are given by the familiar expression:

\be  J_{\mu\nu}=-(a_\mu\p_\nu - a_\nu\p_\mu) 
+S_{\mu\nu}  \ee

For the scalar field, we have $S_{\mu\nu}=0$ and  
for the Proca fields $(S_{\mu\nu})_{\alpha\beta}\break
\equiv(\Sigma_{\mu\nu})_{\alpha\beta}=
(g_{\mu\alpha}g_{\nu\beta}-g_{\mu\beta}g_{\nu\alpha})$. 
For the Dirac field, the $4\times4$ matrices 
$S_{\mu\nu}\equiv\sigma_{\mu\nu}$  
provide a direct sum of the $(\frac12,0)$ 
and $(0,\frac12)$ representations of the restricted Lorentz group. 
Since parity transforms these representations into each other, the 
direct sum is necessary to provide an irreducible representation of 
the complete Lorentz group.  When invariance under parity is not 
required, which is the case of neutrino fields, two-component 
spinors can be used. 

\subsection{The de Sitter higher-order subalgebra} 

The condition (\ref{plorentz1}) is not strong enough to provide 
irreducible representations of the Poincar\'e group. In general, 
reducing the representation will require imposing  
higher-order polarization conditions 
\cite{[higherpjmp],[marmo],[higherpcmp]}. 

It is clear that any (higher-order) polarization must 
contain the Casimir operators of the Poincar\'e group,  
$P^2= \eta^{\mu\nu}P^L_\mu P^L_\nu$ and 
$W^2=\eta^{\mu\nu}W^L_\mu W^L_\nu$, 
with $W_\mu=\frac12\epsilon_{\mu\alpha\beta\nu}
J^{\alpha\beta}P^\nu$. 
Therefore, the functions $\Phi(x)$,  
if supporting an irreducible representation 
of the Poincar\'e group, must verify

\bea P^2\Phi(x)\equiv\square\Phi(x)&=&-m^2\Phi(x)\label{P2}\\
W^2\Phi(x)\equiv \left(-\frac12\square S^{\mu\nu} S_{\mu\nu}
+ S^{\mu\alpha}S_{\mu\beta}\p_\alpha \p^\beta\right)\Phi 
&=&-m^2s(s+1)\Phi(x)\label{W2}\eea 
where $\p_\alpha=\partial/\partial a^{\alpha}$ and so on and 
$\square=\eta^{\mu\nu}\p_\mu\p_\nu$. 
The constant $m$, which we take to be real and non-negative, is 
the mass of the (quanta of) the field. The constant $s$, which we 
take to be discrete (positive integer or half integer) is the 
spin of (the quanta) of the field. 

Let us now consider the following operators 
of the enveloping algebra of the Poincar\'e group 

\be {}^\lambda X_\mu= \lambda P_\mu + P^\rho J_{\rho\nu} \ee 
where the unspecified real number $\lambda$ 
will be determined later. 
These second-order operators ${}^\lambda X_\mu$, together 
with $J_{\mu\nu}$, weakly close a de Sitter (sub-)algebra: 

\bea \left[{}^\lambda X^L_\mu,{}^\lambda X^L_\nu\right]&=&
J^L_{\mu\nu} P^2\nonumber\\
\left[J^L_{\mu\nu},{}^\lambda X^L_\rho\right]&=&
\eta_{\nu\rho}{}^\lambda X^L_\mu -
\eta_{\mu\rho}{}^\lambda X^L_\nu\>\label{algdeSitter}\eea 
Therefore, the corollary in Sect. 2 applies here with $H ={\cal T}$. 
Thus, we may get an irreducible representation of the 
Poincar\'e group over fields on Minkowski space 
by imposing the de Sitter algebra $<J^L_{\mu\nu}, 
{}^\lambda X^L_\nu>$, together with $P^2$ and $W^2$, as  
the (higher-order) polarization.  

We impose (the rest of) the de Sitter 
polarization conditions as follows: 

\be ({}^\lambda X^L_\mu \Psi)^\alpha \equiv  \left((\lambda P^L_\mu + 
{P^L}^\rho J^L_{\rho\nu})\Psi\right)^\alpha  =
im\left(\rho_\mu\right)^\alpha_\beta\Psi^\beta\label{polX} \ee 
where $\rho_\mu$ are a set of matrices which must provide a 
finite-dimensional representation of the de Sitter algebra. 
In particular, we must have 

\bea \left[\rho_\mu,\rho_\nu\right]&=&S_{\mu\nu}\label{desitter40}\\
\left[S_{\mu\nu},\rho_\rho\right]&=&
\eta_{\nu\rho}\rho_\mu -\eta_{\mu\rho}\rho_\nu\label{desitter40b}\eea 

The last equation implies 

\be S(\Lambda)\rho^\mu S(\Lambda)^{-1}={\Lambda^\mu}_\nu \rho^\nu
\label{desitter42}\ee

\noindent which, together with Eqs. (\ref{polX}) and (\ref{plorentz2}),  
implies that the wave functions $\Phi$ obey the equations: 

\be \left(\lambda\frac{\partial}{\partial a^\mu} +  
S_{\mu\nu}\frac{\partial}{\partial a_\nu}\right)\Phi(a)=im\rho_\mu\Phi(a)
\label{eqofmot1}\ee  

Since $P^\mu\, {}^\lambda X_\mu = \lambda P^2$,
Eq. (\ref{eqofmot1}) yields 

\be \left(i\rho^\alpha \frac{\partial}{\partial a^\alpha}+\lambda
m\right)\Phi=0 \label{eqofmot2}\ee 

Equations of motion of this form, with matrices $\rho^\mu$  
fulfilling Eqs. (\ref{desitter40},\ref{desitter40b}) have been
extensively studied in the literature. In particular, 
it was shown \cite{[bhabha]} that, for arbitrary spin $s$,
eq. (\ref{eqofmot2}) leads to an equation of the form: 

\be  (s^2\square +\lambda^2m^2)((s-1)^2\square +
\lambda^2m^2)\cdot\cdot\cdot\Phi=0\ee 
where the last factor is $(\lambda^2m^2)$ for integer spin and
$\left((\frac12)^2\square +\lambda^2m^2\right)$ for half-integer spin. 
Therefore, for the special fields --scalar, Proca and Dirac--  
$\lambda$  is fixed (save for an irrelevant sign) whereas, for 
greater spins, it can have different values: 

\be \lambda =\pm s,\quad \pm(s-1),\>...\> \pm
1\>\>(s\>\>\hbox{integer})\quad \pm\frac12 \>\>(s\>\>\hbox{half
integer})\ee  
This can also be seen by realizing, from Eq. (\ref{eqofmot2}),  
that $\lambda$ must
be an eigenvalue of $\rho^0$, which has the same eigenvalues 
as the spin matrices $iS_{ij},\>i,j=1,2,3$. 

In fact, it is easy to show that for $\lambda=s=\frac12$ 
Eqs. (\ref{eqofmot1}) and (\ref{eqofmot2}) are equi-\break valent to each
other and describe the Dirac particle. The matrices $\rho^\mu$ prove
to be $\rho^\mu =\frac12\gamma^\mu$, with $\gamma^\mu$ (a representation
of) the Dirac matrices. Also, for $\lambda=s=1$,  
Eqs. (\ref{eqofmot1}) and (\ref{eqofmot2}) are equivalent to each other 
and describe the Duffin-Kemmer-Petiau field, which,   
depending on the representation taken for 
the matrices $\rho^\mu\equiv\beta^\mu$ --five dimensional or ten
dimensional--  is equivalent to the scalar field or the Proca field, 
respectively (see Appendix).  

\v4mm 
\noindent {\bf Massless fields}
\v3mm 

In the case of massless fields, for which $m^2=0$, the operators
$J_{\mu\nu}$ and ${}^\lambda X_{\rho}$ generate a Poincar\'e algebra. 
Therefore, in this case the polarization condition to be imposed along
with Eq. (\ref{plorentz1}) is: 

\be {}^\lambda X_{\rho}\Psi\equiv  \left((\lambda P^L_\mu + 
{P^L}^\rho J^L_{\rho\nu})\Psi\right)^\alpha  = 0 \label{massless10}\ee 
which  leads to

\be \left(\lambda\frac{\partial}{\partial a^\mu} + 
S_{\mu\nu}\frac{\partial}{\partial a_\nu}\right)\Phi(a)=0 
\label{massless20}\ee  

Here, we do not need to introduce 
the $\rho^\mu$ matrices, and no equation similar to (\ref{eqofmot2}) 
is reached. As $\lambda$ is an eigenvalue 
of $S^{0i}$ -- for instance $S^{03}$ -- we 
obtain a result similar to that for massive fields: $\lambda$ is 
fixed for the special fields, and equals the helicity, whereas it can
have different values for higher-spin fields.  In fact, for
$\lambda=1$, Eq. (\ref{massless20}) reproduces, depending on the
representation for $S_{\mu\nu}$, the wave equation for
the scalar or Maxwell fields \cite{[chandra]}. For $\lambda=\frac12$, 
Eq. (\ref{massless20}) reproduces, after projecting 
over the chirality eigenfields with $(1\pm\gamma^5)$, 
the equation of motion for neutrinos. 

\section{Discussion} 

Our analysis is not meant 
to be exhaustive but rather to show how modern representation 
techniques serve to give   
the subject of relativistic field equations a new look.  
However, we have been able to give a broad if brief 
view on that subject. Although our emphasis has been placed 
on the special fields --    
those with $s\leq1$, which are by far the most relevant ones, 
as no elementary particle has yet been found with $s> 1$ --  
the possibility is open to make a more detailed analysis  
of the higher-spin fields within this formalism. 
Another interesting extension of the present analysis, which is 
already under way \cite{[second]},  
is towards analysing other relativity 
groups such as the de Sitter-like groups, which describe   
fields in curved space.  

Finally, let us briefly discuss the unitarity of the 
representations we have obtained.  
Since our wave functions depend only
on the coordinates of Minkowski spacetime, and their value on different
times can be obtained from the Cauchy data and the equations of motions, 
any invariant scalar product must be of the form 

\be 
<\Phi'|\Phi>= \int_\Sigma{\d \sigma_\mu J^\mu(\Phi',\Phi)} \ee 
where $ \Sigma $ is an space-like Cauchy hypersurface and $J^\mu$ a
bilinear current, $J^\mu =J^\mu(\Phi',\Phi)=(J^\mu(\Phi,\Phi'))^*$, which,  
in order to give rise to a (space-)time-independent scalar
product, must be divergenceless, $\p_\mu J^\mu=0$. 

Currents which fulfill these requirements are the following ones 
\cite{[config]} 

\be J^\mu =\left\{
\ba{ll}
i(\phi'^*\p^\mu\phi-\p^\mu\phi'^*\phi)\quad\hbox{(scalar)}\\
i({F'^*}^{\mu\nu}A_\nu-A'^*_\nu F^{\mu\nu})\quad\hbox{(Proca and Maxwell)}\\
\bar\psi'\gamma^\mu\psi\quad\hbox{(Dirac)}\\
\bar\varphi'\beta^\mu\varphi\quad\hbox{(Duffin-Kemmer-Petiau)}
\ea
\right.
\ee   

Except for the Dirac field, these products  fail
to be positive definite.  
In general, the representations of the complete 
Poincar\'e group which are induced from a finite-component 
representation of its Lorentz subgroup 
fail to have either a positive definite invariant
product (they are not unitary henceforth) or a Hamiltonian which is
bounded from below.  The Dirac field, for instance, is endowed with a 
well-behaved scalar product but fails to have 
a bounded Hamiltonian. The other fields present the reverse behaviour:
a positive Hamiltonian but a non-definite ``scalar'' product. For the 
Dirac field, the situation is reversed by means of its    
``grassmannization'': the Hamiltonian is made positive
whereas the scalar product becomes indefinite. The 
product is then interpreted,
by analogy with other fields, as the electric charge 
of the (configuration or state of) the field. 
For one or the other reason, none of these wave equations 
gives rise to a well-behaved (single-particle) quantum theory. 
The procedure of second quantization is thus required 
(For a more detailed view of the procedure 
of ``second-quantization'' from a group-theoretic standpoint see 
Refs. \cite{[groupfields],[linfields],[empro],[second]}). 

\newpage

\large
\noindent{\Large {\bf Appendix:}} 
{\large {\bf The Clifford and Kemmer algebras 
and the de Sitter group}
\normalsize
\v3mm
Let $\eta_{\mu\nu}$ be a D-dimensional pseudo-euclidean 
flat metric. Let us assume we are able to 
find matrices $\gamma_\mu, \> \mu=1,...,D$
such that 

\be \{\gamma_\mu,\>\gamma_\nu\}=\gamma_\mu\gamma_\nu
+\gamma_\nu\gamma_\mu =2\eta_{\mu\nu} \label{clifford10}\ee 
and let $\rho_\mu\equiv\frac12\gamma_\mu$ and 

\be S_{\mu\nu}= [\rho_\mu,\rho_\nu]\equiv\rho_\mu\rho_\nu-
\rho_\nu\rho_\mu=\frac14[\gamma_\mu,\>\gamma_\nu]\>  \label{clifford20}\ee 
then the matrices $\rho_\mu$ and $S_{\mu\nu}$ close a de Sitter
algebra: 

\bea  
\left[S_{\mu\nu},S_{\alpha\beta}\right]&=&  
\eta_{\nu\alpha}S_{\mu\beta}-
-\eta_{\mu\alpha}S_{\nu\beta}+\eta_{\mu\beta}S_{\nu\alpha}-
\eta_{\nu\beta}S_{\mu\alpha}\nonumber\\
\left[S_{\mu\nu},\rho_\rho\right]&=&
\eta_{\nu\rho}\rho_\mu -\eta_{\mu\rho}\rho_\nu\label{desitteralgebra}\\
\left[\rho_\mu, \rho_\nu\right]&=&S_{\mu\nu}
\nonumber\eea 
which follows from the following identity: 

\be [[A,B],C]=\{A,\{B,C\}\}-\{B,\{A,C\}\}\nonumber \ee 
 
Therefore, if we are given a representation of the Clifford algebra
(\ref{clifford10}), we automatically obtain a representation of the
de Sitter group. 

As is well known, for $D$ even, there is only one irreducible 
representation of the Clifford algebra whereas, for D odd, there are 
two irreducible unitarily-inequivalent representations which nevertheless 
differ only by a sign.  
In four dimensions, the Clifford algebra is irreducibly represented by 
the Dirac matrices. In three dimensions, it is
represented by the Pauly matrices $\sigma_i$. The fact that $[\sigma_i, 
\sigma_j]=\epsilon_{ijk}\sigma_k$ provides direct proof of  
the well-known result that the 
(pseudo-)orthogonal algebras in four dimensions are equivalent 
to a direct sum of two $A_1$ algebras. In two
dimensions the Clifford algebra 
is represented by $\{\sigma^1,\sigma^2\}$, which close  
a $A_1$ algebra. 
\v2mm

The Kemmer algebra is defined by a set of matrices
$\beta^\mu,\>\mu=1,...D$, which obey the relations

\be \beta^\mu\beta^\lambda\beta^\nu + 
\beta^\nu\beta^\lambda\beta^\mu=\eta^{\mu\lambda}\beta^\nu +
\eta^{\nu\lambda}\beta^\mu\label{kemmer1}\ee 

We now define the matrices $S_{\mu\nu}$ as follows: 

\be S_{\mu\nu}= [\beta_\mu,\>\beta_\nu]\>  \label{kemmer20}\ee 
Then, by using the relation (\ref{kemmer1}) and the Jacobi
identity, it is easy to show that the matrices $\beta_\mu$ 
and $S_{\mu\nu}$ also close a de Sitter algebra.  

In 4 dimensions, the Kemmer algebra has, apart from the trivial one, 
two irreducible representations: a five-dimensional one,  
which corresponds to the scalar field and its four spacetime derivatives,  
and a ten-dimensional other which corresponds to the four components  
of a vector field and the six components of its associated stress 
tensor. 

\v3mm
\noindent{\bf Acknowledgements:} M. N. is grateful to the Spanish 
M.E.C., C.S.I.C. and I.M.A.F.F. for a research contract, and to the
Imperial College, where this paper was written, for its boundless
hospitality. M. Calixto acknowledges the Spanish M.E.C for a F.P.I. grant.

This work was partially supported by the DGICYT. 


\newpage
\begin{thebibliography}{99} 

\bibitem{[Weinberg]} S. Weinberg, ``The quantum theory of fields'', 
Cambridge: Cambridge University Press, 1996. Chapters 2 and 5. 

\bibitem{[marmo]} V. Aldaya, J. Guerrero and G. Marmo,   
``Quantization on a Lie group: higher-order polarizations'', contribution 
to {\it Symmetry and Science X}, Bregenz (Austria) 1997 (to appear in the 
proceedings), hep-th/9710002. 

\bibitem{[higherpjmp]} V. Aldaya, J. Bisquert, R. Loll and
J. Navarro-Salas, {\it J. Math. Phys.} {\bf 33}, 3087  (1992). 

\bibitem{[higherpcmp]} V. Aldaya and J. Navarro-Salas, {\it
Comm. Math. Phys.} {\bf 139}, 433 (1991). 

\bibitem{[corson]} E.M. Corson, ``Introduction to tensors, spinors and
relativistic wave equations: (relation structure)'', Blackie, London
1953. 

\bibitem{[fonda]} L. Fonda and G.C. Ghirardi, ``Symmetry principles in 
quantum physics'', Marcel Dekker, New York, 1970. 

\bibitem{[greiner]} W. Greiner, ``Relativistic quantum mechanics: wave
equations'', Springer-Verlag, 1990. 

\bibitem{[naimark]} M.A. Naimark, ``Linear representations of the
Lorentz group'', Pergamon Press, London 1964. 

\bibitem{[taka]} Y. Takahashi, ``An introduction to field
quantization'', Pergamon Press, 1969. 

\bibitem{[umezawa]} H. Umezawa, ``Quantum Field Theory'', North-Holland 
Publishing Co., Amsterdam, 1956. 

\bibitem{[velo]} G. Velo and A.S. Wightman, ``Invariant wave
equations'', Springer-Verlag, Berlin, 1978. 

\bibitem{[groupfields]} V. Aldaya, J.A. de Azcarraga and 
S. Garc\'\i a, {\it J. Phys.} {\bf A21}, 4265 (1988)

\bibitem{[annals]} V. Aldaya and J.A. de Azcarraga, {\it Ann.  
Phys.} N.Y. {\bf 165}, 484 (1985). 

\bibitem{[bhabha]} H.J. Bhabha, {\it Rev. Mod. Phys.} {\bf 17}, 
200 (1945); {\bf 21}, 451 (1949). 

\bibitem{[chandra]} Harish-Chandra, {\it Proc. Roy. Soc.} {\bf A186}, 
502 (1946). 

\bibitem{[config]} M. Navarro, V. Aldaya and M. Calixto, 
{\it J. Math. Phys.} {\bf 37}, 206 (1996), hep-th/9501085. 

\bibitem{[linfields]} M. Navarro, V. Aldaya and M. Calixto, 
{\it J. Math. Phys.} {\bf 38}, 1454 (1997), hep-th/9612068.

\bibitem{[empro]}  V. Aldaya, M. Calixto and M. Navarro.  
``The electromagnetic and Proca fields 
revisited: a unified quantization',  
{\it Int. J. Mod. Phys.} {\bf A12}, 3609 (1997), hep-th/9609083. 

\bibitem{[second]} M. Calixto, V. Aldaya and M. Navarro, 
``Quantum field theory in curved space from a second quantization 
on a group'', {\it J. Math. Phys.} (Submitted), hep-th/9701180. 

\end{thebibliography}
\end{document}